\documentclass[preprint,12pt]{elsarticle}

\usepackage{amssymb}
\usepackage{amsmath,amssymb,amsfonts}
\usepackage{tikz}
\usepackage{tabularx}

\newtheorem{theorem}{Theorem}
\newtheorem{corollary}{Corollary}
\newtheorem{lemma}{Lemma}
\newdefinition{rmk}{Definition}
\newproof{proof}{Proof}
\newtheorem{remark}{Remark}

\journal{Applied Mathematical Modelling}

\begin{document}

\begin{frontmatter}

\title{Reconstruction of shear force in Atomic Force Microscopy from measured displacement of the cone-shaped cantilever tip}

\author[1]{Alemdar Hasanov\fnref{fn1}}
\ead{alemdar.hasanoglu@gmail.com}
\author[2]{Onur Baysal\fnref{fn2}}
\ead{onur.baysal@um.edu.mt}
\author[3]{Alexandre Kawano\corref{cor1}
\fnref{fn3}}
\ead{akawano@usp.br}

\cortext[cor1]{Corresponding author}
\fntext[fn1]{Department of Mathematics, Kocaeli University, Turkey}
\fntext[fn2]{Department of Mathematics, University of Malta, Malta}
\fntext[fn3]{Escola Polit\'{e}cnica, University of S\~{a}o Paulo, São Paulo 05508900, Brazil}

\address[1]{Department of Mathematics, Kocaeli University, Turkey}
\address[2]{Department of Mathematics, University of Malta, Msida, Malta}
\address[3]{Escola Polit\'{e}cnica, University of S\~{a}o Paulo, S\~{a}o Paulo 05508900, Brazil}

\begin{abstract}
In this paper, a dynamic model of reconstruction of the shear force $g(t)$ in the Atomic Force Microscopy (AFM) cantilever tip-sample interaction is proposed. The interaction of the cone-shaped cantilever tip with the surface of the specimen (sample) is modeled by the damped Euler-Bernoulli beam equation $\rho_A(x)u_{tt}$ $+\mu(x)u_{t}+(r(x)u_{xx}+\kappa(x)u_{xxt})_{xx}=0$, $(x,t)\in (0,\ell)\times (0,T)$, subject to the following initial, $u(x,0)=0$, $u_t(x,0)=0$ and boundary, $u(0,t)=0$, $u_{x}(0,t)=0$, $\left (r(x)u_{xx}(x,t)+\kappa(x)u_{xxt} \right )_{x=\ell}=M(t)$, $\left (-(r(x)u_{xx}+\kappa(x)u_{xxt})_x\right )_{x=\ell}=g(t)$ conditions, where $M(t):=2h\cos \theta\,g(t)/\pi$ is the momentum generated by the transverse shear force $g(t)$. For the reconstruction of $g(t)$ the measured displacement  $\nu(t):=u(\ell,t)$ is used as an additional data. The least square functional $J(F)=\frac{1}{2}\Vert u(\ell,\cdot)-\nu \Vert_{L^2(0,T)}^2$ is introduced and an explicit gradient formula for the Fr\'echet derivative through the solution of the adjoint problem is derived. This allows to construct a gradient based numerical algorithm for the reconstructions of the shear force from noise free as well as from random noisy measured output $\nu (t)$. Computational experiments show that the proposed algorithm is very fast and robust. This allows to develop a numerical "gadget" for computational experiments of generic AFMs.
\end{abstract}

\end{frontmatter}


\section{Introduction}
\label{intro}

\begin{figure}
  \centering
 \hspace*{-.6cm} \begin{tikzpicture}[scale=.86]
      \draw[color=black,ultra thick] (0,-0.6) -- (0,0.5);
      \draw[color=gray] (-0.25,-0.5) -- (0,-0.25);
      \draw[color=gray] (-0.25,-0.25) -- (0,-0);
      \draw[color=gray] (-0.25,0) -- (0,0.25);
      \draw[color=gray] (-0.25,0.25) -- (0,0.5);
      \draw[color=gray] (-0.25,0.5) -- (0,0.75);
      \draw[dashed] (0,0) -- (10,0);
       \draw[dashed] (8.5,-0.8) -- (9.9,-0.8);
       \draw[color=gray] (5.5,-0.8) -- (8.5,-0.8);
       \draw[thin, ->] (5.8,-0.5)-- (5.8,-0.18);
       \draw[thin, ->] (5.8,-0.5)-- (5.8,-0.78);
       \node[label=right:{\scriptsize $h$-tip length}] at (5.6,-0.5) {};
      \draw[color=black,thick] (0,0.15) -- (10,0.15);
      \draw[color=black,thick] (0,-0.15) -- (10,-0.15);
      \draw[color=gray] (10,-0.15) -- (10,0.15);
       \draw[color=black,thick] (8.5,-0.55) -- (11.0,-0.55);
      \draw[color=black,thick] (10.0,-0.15) -- (10.0,0.15);
       \draw[color=black,thick] (8.5,-1.45) -- (11.0,-1.45);
      \draw[color=black,thick] (9.6,-0.17) -- (9.8,-0.78);
      \draw[color=black,thick] (10.0,-0.17) -- (9.8,-0.78);
       \draw[color=black,thick] (8.5,-1.45) -- (8.5,-0.55);
       \draw[color=black,thick] (11.0,-1.45) -- (11.0,-0.55);
      \draw[thick,->] (10,0) -- (12,0) node [right]{$x$};
      \draw[thick,->] (0,-1.5) -- (0,1.75) node [left]{$u$};
      \node[label=right:{\scriptsize sample}] at (6.8,-1.3) {};
      \draw[thin, ->] (7.7,-1.0)-- (8.4,-1.0);
      \node[label=right:{\scriptsize$ u(0,t)=0$}] at (-.25,-0.45) {};
      \node[label=right:{\scriptsize$ u_{x}(0,t)=0$}] at (-.25,-0.85) {};
      \node[label=left:{\scriptsize $\left(r(x)u_{xx}+\kappa(x)u_{xxt}\right)\vert_{x=\ell}=M(t)$}] at (4.1*3.14,0.95) {};
      \node[label=left:{\scriptsize$-\left(r(x)u_{xx}+\kappa(x)u_{xxt}\right)_x\vert_{x=\ell}=g(t)$}] at (4.1*3.14,0.5) {};
      \node[label=right:{\scriptsize $M(t):=\,\frac{2h\cos \theta}{\pi}\, g(t)$}] at (9.84,-1.02) {};
       \node[label=right:{\scriptsize $g(t)$}] at (8.8,-1.15) {};
       \node[label=right:{\small$\ell$}] at (9.7,-0.23) {};
       \draw[thick, ->] (9.8,-1.3)-- (9.8,-0.82);
       \draw[thick, ->] (10.3,-0.8)-- (9.9,-0.8);
        \end{tikzpicture}
  \caption{Schematic diagram of AFM cone-shaped cantilever tip-sample interaction} \label{fig-1}
  \end{figure}
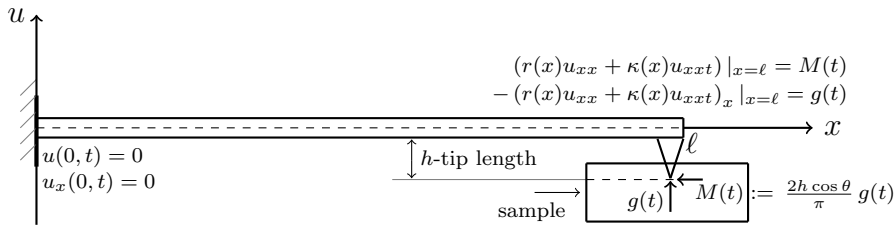

Micro-cantilever plays a key role in nanomachining process using an AFM which was originally developed to provide surface topography information \cite{Binnig:1986}. Nowadays, AFM can provide high resolution images in different settings including ambient, aqueous and vacuum environments. In standard AFMs, the micro-cantilever is mounted horizontally and the devices are operated in a contact or intermittent-contact mode (Fig. \ref{fig-1}).  The cantilever tip-sample interaction creates a transverse shear force and a bending moment on the tip of the cantilever \cite{AFM2012}. Estimation of the unknown shear force signal allows better interpretation and understanding of  scan results. Since this force can only be measured indirectly, via a laser based sensor system, various models and inversion algorithms was developed for reconstruction of the transverse shear force in atomic and dynamic force microscopy, through the measured cantilever tip deflection (see \cite{Antognozzi:2001, Chang:2004, Zhang:2022} and references therein).

For the AFM cone-shaped cantilever tip-sample interaction, a simple mathematical model for the shear force reconstruction problem has first been proposed in \cite{Chang:2004}, within the Euler-Bernoulli beam theory. Namely, the model considers the cutting system as the inverse problem of reconstructing the cutting force $F_y(t)$ in
\begin{eqnarray}\label{c-1}
\left\{ \begin{array}{ll}
y_{xxxx}+\frac{\rho A}{EI}\, y_{tt}=0, ~x \in (0,L ), \\ [1pt]
y(x,0)=y_{t}(x,0)=0, ~x \in (0, L), \\ [1pt]
y(0,t)=y_x(0,t)=0,\\ [1pt]
y_{xx}(L,t)=\frac{-F_x(t)h}{EI},~y_{xxx}(L,t)=\frac{F_y(t)}{EI},~t \in [0,T],
\end{array} \right.
\end{eqnarray}
from the measured displacement
\begin{eqnarray} \label{c-2}
Y (L, t) :=  y(L,t),~ t \in [0,T].
\end{eqnarray}
using available displacement measurement. Here, $F_x = \left (2 h\cos \theta \,F_y\right)/\pi$ for a cone-shaped cantilever with the half-conic angle $\theta$, and $h>0$ is the cantilever tip length. This is an inverse problem with two Neumann inputs. Note that a similar inverse problem with one Neumann input was proposed in \cite{Hasanov-Baysal-Sebu:2019}. A detailed analysis of inverse problems of identifying the unknown transverse shear force in the Euler–Bernoulli beam with Kelvin–Voigt damping was given in \cite{Kumarasamy-Hasanov:2023}.

It is important to emphasize that in the AFM cone-shaped cantilever tip-sample interaction model (\ref{c-1}) is based on the simplified and constant coefficient Euler-Bernoulli beam equation, without the viscous external ($\mu(x) u_t$) and the internal or Kelvin-Voigt ($\left (\kappa (x) u_{xxt}\right )_{xx}$) damping terms. Thus, in these models, not all physical properties of the cantilever are  taken into account. However, the influence of these above mentioned properties on the dynamic behavior of the AFM cantilever is enormous, and needs to be studied carefully \cite{Banks:1991, Shen:2004}.

In this paper we propose a mathematical model of tip-sample processing in AFM with two Neumann inputs. This model is a generalization of existing mathematical models in the sense that;\\
(a) the Euler-Bernoulli equation contains all the physical variable coefficients, including the both damping terms;\\
(b) the time interval during which it is necessary to produce an experimental data, i.e. measured output, can be small enough;\\
(c) the measured output contains random noise;\\
(d) the inputs in the model may not be smooth enough.

Within the proposed model, we formulate the inverse problem of reconstructing the unknown shear force from measured displacement of the cone-shaped cantilever tip.
We provide a detailed mathematical and numerical analysis of the problem. Based on this analysis, we derive an explicit gradient formula for the least square functional. This allows us to construct an effective and fast reconstruction algorithm, as the presented results of  computational experiments show.

\section{Vibration model of tip-sample processing: the reconstruction problem}

The sample processing with AFM cone-shaped cantilever, shown schematically in Fig.  \ref{fig-1}, is modeled as a damped Euler-Bernoulli beam. This cantilever, with length $\ell>0$ and cross-sectional area $A_s(x)$, is clamped at the left end $x=0$. The tip-sample contact is modeled by a vertical reaction force, which is the transverse shear force with the negative sign, that is  $-g(t)$, and the moment $M(t):=-2h\cos \theta\,g(t)/\pi$, generated by this force, where $h,\,\theta>0$ are the tip length and half-conic angle, respectively. Then the sample processing vibration model is governed by the following initial boundary value problem for the damped Euler-Bernoulli equation:
\begin{eqnarray}\label{1-1}
\left\{ \begin{array}{ll}
\rho_A(x) u_{tt}+\mu(x)u_{t}+ (r(x)u_{xx}+\kappa(x)u_{xxt})_{xx} =0,\, (x,t)\in \Omega_{T},\\ [5pt]
u(x,0)=u_{t}(x,0)=0, ~x \in (0,\ell ), \\ [5pt]
u(0,t)=u_x(0,t)=0,~ \left (r(x)u_{xx}+\kappa(x)u_{xxt}\right)_{x=\ell}=M(t),\\ [2pt]
\qquad \qquad \qquad \left (-(r(x)u_{xx}+\kappa(x)u_{xxt})_x \right )_{x=\ell}=g(t),~t \in [0,T],
\end{array} \right.
\end{eqnarray}
where $\Omega_T:=(0,\ell)\times(0,T)$, and the final time instance $T>0$ may be small enough. Here and below, $\rho_A(x):=\rho(x)A_s(x)$, while $\rho(x)>0$ and $A_s(x)>0$ are the mass density and the cross-sectional area of the nonhomogeneous cantilever, $r(x):=E(x)I(x)>0$ is the flexural rigidity (or bending stiffness) of the cantilever while $E(x)>0$ is the elasticity modulus and $I(x)>0$ is the moment of inertia. The coefficient $\kappa(x):=c_d(x)I(x)$ represents energy dissipated by friction internal to the beam, while $c_d>0$ is the strain-rate damping coefficient \cite{Banks:1991}. The external and internal damping mechanisms are given by the terms $\mu(x)u_t$ and $(\kappa(x)u_{xxt})_{xx}$, respectively. The coefficients $\mu(x)\ge 0$ and $\kappa(x)>0$ are called the viscous (internal) damping and the strain-rate or Kelvin-Voigt damping coefficients, respectively.

The transverse shear force $g(t)$ in (\ref{1-1}) is assumed to be unknown and needs to be determined from knowledge of the measured displacement $\nu(t)$ of the cone-shaped tip:
\begin{eqnarray} \label{1-2}
\nu(t):=  u(\ell,t),~ t \in [0,T].
\end{eqnarray}

Thus the \emph{inverse boundary value problem} here is to reconstruct the unknown transverse shear force $g(t)$ in (\ref{1-1}) from knowledge of the measured displacement $\nu(t)$ defined in (\ref{1-2}).

As noted above, the model governed by (\ref{1-1}) and (\ref{1-2}) describes an inverse problem with two inputs $M(t)$ and $g(t)$. As we shall see, this results in a number of differences and additional problems, unlike the single-input inverse problems considered in \cite{Hasanov-Baysal-Sebu:2019, Kumarasamy-Hasanov:2023}. Note also that due to the conical geometry of the rod, the relationship $M(t):=\left(2h\cos \theta \, g(t)\right )/\pi$ is defined between the inputs.

We assume that the following basic conditions are satisfied:
\begin{eqnarray} \label{1-3}
\left \{ \begin{array}{ll}
\rho_A, \mu, r, \kappa \in L^\infty(0,\ell),\\ [3pt]
g \in  H^1(0,T),g(0)=0,\\ [3pt]
0<\rho_0\leq \rho_A(x)\leq \rho_1,~0\leq \mu_0\leq \mu(x)\leq \mu_1,\\ [3pt]
0<r_0\leq r(x)\leq r_1,~0<\kappa_0\le \kappa(x)\le \kappa_1,\,  x\in (0,\ell).
\end{array} \right.
\end{eqnarray}

Introduce the set of admissible shear forces
\begin{eqnarray} \label{1-4}
\mathcal{G}:=\{g \in H^1(0,T)\,:\, g(0)=0,~\Vert g \Vert_{H^1(0,T)}\le C_g,\},
\end{eqnarray}
where $C_g>0$ is a constant independent on $g(t)$. Denote by $u(x,t;g)$ the solution of the forward problem (\ref{1-1}) for a given $g \in \mathcal{G}$, while $u(\ell,t;g)$ in defined as an \emph{output}. Introduce the Neumann-to-Dirichlet operator:
\begin{eqnarray} \label{1-5}
\left \{ \begin{array}{ll}
(\Psi g)(t):=u(\ell,t;g), ~ t \in [0,T], \\ [2pt]
		\Psi :\mathcal{G} \subset H^1(0,T)\mapsto L^2(0,T),
\end{array} \right.
\end{eqnarray}
defined on the set of admissible shear forces. In view of this operator, we can reformulate the inverse problem as the linear operator equation:
\begin{eqnarray} \label{1-6}
u(\ell,t;g)=\nu(t),~ t \in [0,T]
\end{eqnarray}
Since the \emph{measured output} $\nu(t)$ obtained as a result of measurement, it contains random noise. Hence the exact equality between the output $u(\ell,t;g)$ and the measured outputs $\nu(t)$ can never be achieved. As a consequence, there can never be an exact solution to the inverse problem (\ref{1-1})-(\ref{1-2}).

We introduce the Tikhonov functional
\begin{eqnarray}\label{1-7}
J(g):=\frac{1}{2}\,\Vert u(\ell,\cdot;g)-\nu \Vert_{L^2(0,T)}^2,~g \in \mathcal{G},~
\nu \in L^2(0,T)
\end{eqnarray}
and look for the quasi-solution of the inverse problem (\ref{1-1})-(\ref{1-2}): \emph{Find $g \in \mathcal{G}$ such that}
\begin{eqnarray}\label{1-8}
J(g)=\inf_{\tilde{g} \in \mathcal{G}}\, J(\tilde{g}).
\end{eqnarray}

\section{Necessary estimates for the weak solution of problem (\ref{1-1})}

In the case when $M(t)=0$, the existence and uniqueness of the weak solution
$u\in L^2(0,T; \mathcal{V}^2(0,\ell))$, with $u_t\in L^2(0,T;L^2(0,\ell))$ and $u_{tt}\in L^2(0,T;H^{-2}(0,\ell))$ of the initial boundary value problem (\ref{1-1}) is proved in \cite{Kumarasamy-Hasanov:2023}, where
\begin{eqnarray*}
\mathcal{V}^2(0,\ell):=\{v\in H^2(0,\ell):\, v(0)=v(\ell)=0\}.
\end{eqnarray*}
For the direct problem (\ref{1-1}) the same results can be proved in the same way. We derive here some a priori estimates for the weak solution which are necessary in the analysis of the inverse problem (\ref{1-1})-(\ref{1-2}).

\medskip
\begin{theorem}\label{Theorem-1}
Assume that the inputs in (\ref{1-1}) satisfy the basic conditions (\ref{1-3}).
Then the following estimates holds:
\begin{eqnarray}\label{11}
\left. \begin{array}{ll}
\Vert u_{xx} \Vert^2_{L^{\infty}(0,T;L^2(0,\ell))} \leq C^2_1\, \Vert g' \Vert^2_{L^2(0,T)},\\ [9pt]
\Vert u_{xx} \Vert^2_{L^2(0,T;L^2(0,\ell))} \leq C^2_2\, \Vert g' \Vert^2_{L^2(0,T)},\\ [9pt]
\displaystyle \Vert u_{t} \Vert^2_{L^2(0,T;L^2(0,\ell))} \leq \frac{r_0}{2 \rho_0}\,C^2_2\, \Vert g' \Vert^2_{L^2(0,T)},\\ [11pt]
\displaystyle \Vert u_{xxt} \Vert^2_{L^2(0,T;L^2(0,\ell))} \leq \frac{r_0}{4\kappa_0}\,C^2_2\, \Vert g' \Vert^2_{L^2(0,T)},
\end{array} \right.
\end{eqnarray}
where
\begin{eqnarray}\label{12}
\left. \begin{array}{ll}
C^2_1=C^2_0\, \left (1+ C^2_{\theta}\right )\,C^2_e,~C^2_2=C^2_0\,\left (1+ C^2_{\theta}\right )\, \left (1+C^2_e\right ), ~C^2_e=\exp(T), \\ [7pt]
\displaystyle C^2_0=\frac{4\,\widehat{\ell}\,(1+T)}{r_0^2},~ C^2_{\theta}=\left(\frac{2h\cos \theta}{\pi}\right)^2,~\widehat{\ell}=\ell+\ell^3/3,~
\end{array} \right.
\end{eqnarray}
and $r_0, \rho_0,  \kappa_0>0 $ are the constants introduced in (\ref{1-3})
\end{theorem}
{\bf Proof.} Multiply both sides of equation (\ref{1-1}) by $2 u_t(x,t)$, integrate it over $\Omega_t:=(0,\ell)\times (0,t)$, $t\in (0,T]$, and employ the identities
\begin{eqnarray*}
\left. \begin{array}{ll}
\displaystyle 2\,\int_0^t\int_0^\ell (r(x)u_{xx})_{xx} u_{\tau} dx d\tau= 2\, \int_0^t\int_0^\ell [(r(x)u_{xx})_x u_{\tau}-r(x)u_{xx} u_{x\tau}]_x dx d\tau  \nonumber \\ [8pt]
\qquad \qquad \qquad \qquad \qquad \qquad \qquad  \displaystyle + \,\int_0^t\int_0^\ell \left (r(x)u_{xx}^2\right )_{\tau} dx d\tau,~t \in (0,T], \nonumber \\ [11pt]
\displaystyle 2\,\int_0^t\int_0^\ell (\kappa(x)u_{xx\tau})_{xx} u_{\tau} dx d\tau= 2\, \int_0^t\int_0^\ell [(\kappa(x)u_{xx\tau})_x u_{\tau}-\kappa(x)u_{xx\tau} u_{x\tau}]_x dx d\tau  \nonumber \\ [8pt]
\qquad \qquad \qquad \qquad \qquad \qquad  \qquad \displaystyle + \,\int_0^t\int_0^\ell \kappa(x)\left (u_{xx\tau}\right )^2 dx d\tau,~t \in (0,T].
 \end{array} \right.
\end{eqnarray*}
Applying the integration by parts formula multiple times, using the initial and boundary conditions in (\ref{1-1}) we obtain the following energy identity:
\begin{eqnarray*}
\int_0^\ell \left [\rho_A(x) u_t^2+r(x)u^2_{xx}\right ]dx +2 \int_0^t \int_0^\ell \mu(x)u_{\tau}^2dx\,d\tau \qquad \qquad\qquad \qquad \quad \\ [1pt]
\qquad  + 2 \int_0^t \int_0^\ell \kappa(x)u_{xx\tau}^2dx\,d\tau =2\int_0^t M(\tau) u_{x\tau}(\ell,\tau)d \tau+2\int_0^t g(\tau)u_{\tau}(\ell,\tau)d \tau,
\end{eqnarray*}
for all $t \in (0,T]$. Now, applying the integration by parts to the right hand side integrals and the use the $\varepsilon$-inequality, we have:
\begin{eqnarray*}
2\int_0^t M(\tau) u_{x\tau}(\ell,\tau)d \tau+2\int_0^t g(\tau)u_{\tau}(\ell,\tau)d \tau \qquad \qquad\qquad \qquad \qquad \qquad   \\ [1pt]
\qquad \le \varepsilon\, \left [u_x^2(\ell,t)+u^2(\ell,t)+ \int_0^t u_x^2(\ell,\tau)d\tau +
\int_0^t u^2(\ell,\tau)d\tau \right ] \qquad  \qquad \quad \\ [1pt]
\qquad \qquad  + \frac{1}{\varepsilon}\, \left [M^2(t)+g^2(t)+ \int_0^t \left (M'(\tau)\right)^2d\tau +\int_0^t \left (g'(\tau)\right)^2d\tau \right ],\,~t \in (0,T].
\end{eqnarray*}
Use also the auxiliary inequalities (\cite{Hasanov-Romanov:2021}, Ch. 11):
\begin{eqnarray}
		u^2(\ell,t)\le \frac {\ell^3}{3}  \int_0^\ell u_{xx}^2(x,t) dx,~u_x^2(\ell,t)\le \ell\,  \int_0^\ell u_{xx}^2(x,t) dx,\,u\in \mathcal{V}^2(0,\ell), 	\label{13} \\[1pt]
		M^2(t) \le T\, \Vert M^{\prime}\Vert^2_{L^2(0,T)}, ~g^2(t) \le T\, \Vert g^{\prime}\Vert^2_{L^2(0,T)},~~g\in H^1(0,T), ~g(0)=0,\nonumber
	\end{eqnarray}
for all $t \in [0,T]$. Taking these inequalities with the inequality
\begin{eqnarray*}
	 \Vert M\Vert^2_{L^2(0,T)}\le C^2_{\theta}\, \Vert g\Vert^2_{L^2(0,T)},
	\end{eqnarray*}
into account in the energy identity above, we get the following inequality:
\begin{eqnarray*}
\rho_0\,\int_0^\ell u_t^2dx +\left (r_0-\widehat{\ell}\,\varepsilon\right )\,\int_0^\ell u^2_{xx}dx +2 \int_0^t \int_0^\ell \mu(x)u_{\tau}^2dx\,d\tau \qquad\qquad \\ [1pt]
\qquad  + 2 \int_0^t \int_0^\ell \kappa(x)u_{xx\tau}^2dx\,d\tau \le \varepsilon\, \int_0^t  \int_0^\ell u_{xx}^2dx\,d\tau \\ [1.5pt]
+\displaystyle \frac{1+T}{\varepsilon}\,(1+C^2_{\theta})\, \int_0^T \left (g'(t) \right )^2dt,~t \in [0,T],
\end{eqnarray*}
where $\widehat{\ell},C_{\theta}>0$ are the constants introduced in (\ref{12}). Choosing here the arbitrary parameter $\varepsilon>0$ from the condition  $r_0-\widehat{\ell}\,\varepsilon>0$ as $\varepsilon=r_0/(2\widehat{\ell})$ we finally obtain
the main integral inequality:
\begin{eqnarray}\label{14}
\displaystyle \rho_0\,\int_0^\ell u_t^2dx +\frac{r_0}{2}\,\int_0^\ell u^2_{xx}dx +2 \int_0^t \int_0^\ell \mu(x)u_{\tau}^2dx\,d\tau \qquad\qquad \nonumber \\ [1pt]
\qquad  + 2 \int_0^t \int_0^\ell \kappa(x)u_{xx\tau}^2dx\,d\tau \le \frac{r_0}{2}\, \int_0^t  \int_0^\ell u_{xx}^2dx\,d\tau \nonumber \\ [1.5pt]
+\displaystyle \frac{r_0}{2}\,C_0^2\,\left (1+C^2_{\theta}\right )\, \int_0^T \left (g'(t) \right )^2dt,~t \in [0,T],
\end{eqnarray}
where $C_0>0$ is the constant introduced in (\ref{12}).

The first consequence of (\ref{14}) is the  inequality
\begin{eqnarray*}
\int_0^\ell u^2_{xx}dx  \le \int_0^t  \int_0^\ell u_{xx}^2dx\,d\tau +C_0^2\,\left (1+C^2_{\theta}\right )\, \int_0^T \left (g'(t) \right )^2dt,~t \in [0,T].
\end{eqnarray*}
With the Gr\"onwall-Bellmann inequality this implies:
\begin{eqnarray}\label{15}
\int_0^\ell u^2_{xx}dx  \le C_0^2\,\left (1+C^2_{\theta}\right )\,  \Vert g' \Vert^2_{L^2(0,T)} \exp(t),~t \in [0,T].
\end{eqnarray}
Both of the first two estimates in (\ref{11}) are easily derived from this inequality.

The second consequence of (\ref{14}) is the  inequality
\begin{eqnarray*}
\displaystyle \rho_0\,\int_0^\ell u_t^2dx \le \frac{r_0}{2}\,\int_0^t  \int_0^\ell u_{xx}^2dx\,d\tau +\frac{r_0}{2}\, C_0^2\,\left (1+C^2_{\theta}\right ) \int_0^T \left (g'(t) \right )^2dt,~t \in [0,T].
\end{eqnarray*}
With (\ref{15}) this leads to the third estimate in (\ref{11}).

The fourth estimate in (\ref{11}) is proved in the same way.
\hfill$\Box$

\medskip
\begin{remark}\label{Remark-1}
The results of Theorem \ref{Theorem-1} are valid, with slightly different from the constants introduced in (\ref{12}), also for the case where the consistency  condition $g(0)=0$ in (\ref{1-3}) is not met.
\end{remark}

\begin{corollary}\label{Corollary-1}
Assume that conditions of Theorem \ref{Theorem-1} hold.
Then for the $H^1$-norm of the output $u(\ell,t;g)$ the following trace estimate holds:
\begin{eqnarray}\label{16}
\displaystyle \Vert u(\ell,\cdot;g) \Vert^2_{H^1(0,T)}\le C^2_3 \Vert g' \Vert^2_{L^2(0,T)},~
C^2_3=\frac {\ell^3}{3} \left (C^2_1+\frac {r_0}{4\kappa_0}\,C^2_2\right).
\end{eqnarray}
\end{corollary}
Proof follows from the trace inequalities
\begin{eqnarray*}
\displaystyle \Vert u(\ell,\cdot;g) \Vert^2_{L^2(0,T)}\le
		\frac {\ell^3}{3}\, C_1^2 \Vert g' \Vert^2_{L^2(0,T)}, \\[1pt]
\displaystyle \Vert u_t(\ell,\cdot;g) \Vert^2_{L^2(0,T)}\le
		\frac {\ell^3}{3}\,\frac {r_0}{4\kappa_0}\, C_2^2 \Vert g' \Vert^2_{L^2(0,T)},
	\end{eqnarray*}
which are the consequence of the first inequality in (\ref{13}) and estimates in (\ref{11}). \hfill$\Box$

\section{Analysis of the inverse problem}

The compactness property is one of the main properties of the input-output operators corresponding to problems, since the ill-posedness of an inverse problem is the result of this property. For the simplified version, with one Neumann input ($M(t)=0$) and with $\kappa(t)=0$, the compactness of the Neumann-to-Dirichlet operator (\ref{1-5}) is proven in \cite{Hasanov-Baysal-Sebu:2019} for the regular weak solution. For the model we are considering, the regularity condition is not necessary, as we shall see below. That is, this property is also preserved in the case of the weak solution, which shows the role of the Kelvin–Voigt damping coefficient $\kappa(x)>0$.

\medskip
\begin{lemma}\label{Lemma-1}
Under the basic conditions (\ref{1-3}), the Neumann-to-Dirichlet operator
$\Psi :\mathcal{G} \subset H^1(0,T)\mapsto L^2(0,T)$ introduced in (\ref{1-5}) is a linear compact operator.
\end{lemma}
{\bf Proof.} Let $ \{g_m\} \subset \mathcal{G}$, $m=1,2,\,...\,$, be a sequence of inputs, bounded in the norm of $H^1(0,T)$, according to the definition (\ref{1-4}) of set of admissible shear forces. Denote by $ \{u^{(m)}(x,t)\}$, where $u^{(m)}(x,t):=u(x,t;g_m)$, the corresponding sequence of weak solutions  of the direct problem (\ref{1-1}). By the estimate (\ref{16}), the sequence of outputs $ \{u^{(m)}(x,t)\}$ is bounded in $H^1(0,T)$. Then by the Rellich-Kondrachov compactness theorem, $\psi$ is compact operator.

 \hfill$\Box$

\medskip
\begin{lemma}\label{Lemma-2}
Assume that the basic conditions (\ref{1-3}) hold. Then the Neumann-to-Dirichlet operator is Lipschitz continuous, that is
\begin{eqnarray}\label{17}
\Vert \Phi g_1 - \Phi g_2\Vert_{L^2(0,T)} \leq L_0 \Vert g'_1- g'_2\Vert_{L^2(0,T)},~\mbox{for all}~g_1,g_2 \in \mathcal{G},
\end{eqnarray}
with here $L_0=\sqrt{\ell^3/3}\,C_1>0$ is the Lipschitz constant and $C_1>0$ is the constant introduced in (\ref{12}).
\end{lemma}
{\bf Proof.} Let $u_k(x,t):=u(x,t;g_k)$, $k=1,2$, be two weak solutions of the direct problem (\ref{1-1}) corresponding to the inputs  $g_1,\,g_2 \in \mathcal{G}$. Then the function  $\delta u(x,t)=u_1(x,t)-u_2(x,t)$ solves the problem
\begin{eqnarray}\label{18}
\left\{ \begin{array}{ll}
\rho_A(x) \delta u_{tt}+\mu(x) \delta u_{t}+ (r(x)\delta u_{xx}+\kappa(x)\delta u_{xxt})_{xx} =0,\, (x,t)\in \Omega_{T},\\ [5pt]
\delta u(x,0)=\delta u_{t}(x,0)=0, ~x \in (0,\ell ), \\ [5pt]
\delta u(0,t)=\delta u_x(0,t)=0,~ \left (r(x)\delta u_{xx}+\kappa(x)\delta u_{xxt}\right)_{x=\ell}=\delta M(t),\\[2pt]
\qquad \qquad \qquad \left (-(r(x) \delta u_{xx}+\kappa(x) \delta u_{xxt})_x \right )_{x=\ell}=\delta g(t),~t \in [0,T],
\end{array} \right.
\end{eqnarray}
subject to the inputs $\delta M(t)=C^2_{\theta} \,\delta g(t)$ and $\delta g(t)=g_1(t)- g_2(t)$. By the definition (\ref{1-5}) of the input-output operator we have:
\begin{eqnarray*}
\Vert \Phi g_1 - \Phi g_2 \Vert^2_{L^2(0,T)}=\Vert \delta u(\ell,\cdot)\Vert^2_{L^2(0,T)}.
\end{eqnarray*}
In view of the first inequality in (\ref{13}) and the second estimate in (\ref{11}) applied to the weak solution $\delta u(x,t)$ of problem (\ref{18}) we deduce that
\begin{eqnarray}\label{19}
\Vert \delta u(\ell,\cdot)\Vert^2_{L^2(0,T)} \le\frac {\ell^3}{3}\,C^2_1\, \Vert \delta g' \Vert^2_{L^2(0,T)}.
\end{eqnarray}
This leads to (\ref{17}).
 \hfill$\Box$

The Lipschitz continuity of the Neumann-to-Dirichlet operator leads to the Lipschitz continuity of the Tikhonov functional introduced in (\ref{1-7}), and this, in turn, leads to the existence of the quasi-solution of the inverse problem (\ref{1-1})-(\ref{1-2}), by Theorem 6.5.2 \cite{Hasanov-Romanov:2021}.

\medskip
\begin{theorem}\label{Theorem-2}
Assume that the inputs in (\ref{1-1}) satisfy the basic conditions (\ref{1-3}). Suppose that the measured output $\nu(t)$ belongs to $L^2(0,T)$. Then there exists a quasi-solution of the
inverse problem (\ref{1-1})-(\ref{1-2}) in the set of admissible shear forces $\mathcal{G}$.
\end{theorem}

\section{Fr\'echet differentiability of the Tikhonov functional and gradient formula}

For $g, g+\delta g \in \mathcal{G}$ we find the increment $\delta J(g):=J(g+\delta g)-J(g)$ of the Tikhonov functional introduced in (\ref{1-7}) is
\begin{eqnarray}\label{20}
\delta J(g)=\int_0^\ell \left [u(\ell,t;g) - \nu(t)\right ]\delta u(\ell,t)dt +\frac{1}{2}\int_0^\ell \left (\delta u(\ell,t) \right )^2 dt,
\end{eqnarray}
where $\delta u(x,t)$ is the solution of the sensitivity  problem (\ref{18}).

Multiplying both sides of  equation (\ref{18}) by arbitrary function $\phi(x,t)$, integrating it over $(0,T)$ and applying the integration by parts formula multiple times, we obtain:
\begin{eqnarray}\label{21}
\int_0^T \int_0^{\ell} \left [\rho_A(x)\phi_{tt}-\mu(x)\phi_{t}+\left (r(x)\phi_{xx}-\kappa(x)\phi_{xxt}\right)_{xx}\right] \delta u \,dx dt \qquad \qquad \nonumber \\
\quad + \int_0^\ell \left [\rho_A(x) \delta u_t \phi-\rho_A(x) \delta u \phi_t +\mu(x) \delta u \phi
+\kappa(x)\delta u_{xx}\phi_{xx}\right ]_{t=0}^{t=T} \,dx \quad \nonumber \\
 + \int_0^T \left [\left(r(x) \delta u_{xx}\right)_{x}\phi-r(x)\delta u_{xx} \phi_{x}+
r(x)\delta u_{x} \phi_{xx}-\delta u \left (r(x)\phi_{xx}\right)_{x} \right ]_{x=0}^{x=\ell} \,dt \nonumber \\
+ \int_0^T \left [\left(\kappa(x) \delta u_{xxt}\right)_{x}\phi-\kappa(x)\delta u_{xxt} \phi_{x}-
\kappa(x)\phi_{xxt}\delta u_x \right. \qquad \qquad \qquad \nonumber \\
\left. +\left (\kappa(x)\phi_{xxt}\right)_x \delta u \right ]_{x=0}^{x=\ell} \,dt =0.\qquad
\end{eqnarray}
We require now $\phi(x,t)$ solves the well-posed backward problem
\begin{eqnarray}\label{22}
\left\{\begin{array}{ll}
\rho_A(x) \phi_{tt}-\mu(x)\phi_{t}+ (r(x)\phi_{xx}-\kappa(x)\phi_{xxt})_{xx} =0,\, (x,t)\in \Omega_{T},\\ [1pt]
\phi(x,T)=0,~\phi_t(x,T)=0, ~x \in (0,\ell ), \\ [1pt]
\phi(0,t)=\phi_x(0,t)=0,~ \left (r(x)\phi_{xx}-\kappa(x)\phi_{xxt}\right)_{x=\ell}=0,\\
\qquad \quad \qquad \qquad \left (-(r(x)\phi_{xx}-\kappa(x)\phi_{xxt})_x \right )_{x=\ell}=p(t),~t \in [0,T].
\end{array}\right.
\end{eqnarray}
The control function $p(t)$ here is the arbitrary Neumann input and is specified below.

In view of the initial, final and boundary conditions in (\ref{1-1})  and (\ref{18}) we deduce from (\ref{22}) the following integral relationship:
\begin{eqnarray}\label{23}
\int_0^T p(t) \delta u(\ell,t) dt =\int_0^T \left [ \phi(\ell,t)+C^2_{\theta}\,\phi_x(\ell,t)\right ]\delta g(t) dt,
	\end{eqnarray}
where $C_{\theta}>0$ is the constant introduced in (\ref{12}).

Taking into account the increment formula (\ref{20}) we choose the control function $p(t)$  as follows:
\begin{eqnarray}\label{24}
p(t) =u(\ell,t;g) - \nu(t),~t \in [0,T].
	\end{eqnarray}
The backward problem with this input, i.e. the problem
\begin{eqnarray}\label{25}
\left\{\begin{array}{ll}
\rho_A(x) \phi_{tt}-\mu(x)\phi_{t}+ (r(x)\phi_{xx}-\kappa(x)\phi_{xxt})_{xx} =0,\, (x,t)\in \Omega_{T},\\ [1pt]
\phi(x,T)=0,~\phi_t(x,T)=0, ~x \in (0,\ell ), \\ [1pt]
\phi(0,t)=\phi_x(0,t)=0,~ \left (r(x)\phi_{xx}-\kappa(x)\phi_{xxt}\right)_{x=\ell}=0,\\
\qquad \qquad \left (-(r(x)\phi_{xx}-\kappa(x)\phi_{xxt})_x \right )_{x=\ell}=u(\ell,t;g) - \nu(t),~t \in [0,T],
\end{array}\right.
\end{eqnarray}
is called the adjoint problem corresponding to the inverse problem (\ref{1-1})-(\ref{1-2}).

Substituting (\ref{24}) into (\ref{23}) we obtain the \emph{input-output relationship}:
\begin{eqnarray}\label{26}
\int_0^T \left [u(\ell,t;g) - \nu(t) \right ] \delta u(\ell,t) dt =\int_0^T \left [ \phi(\ell,t)+C^2_{\theta}\,\phi_x(\ell,t)\right ]\delta g(t) dt,
	\end{eqnarray}
which contains the output $u(\ell,t;g)$ and the measured output $\nu(t)$. Comparing (\ref{20}) and (\ref{26}) we deduce that
\begin{eqnarray}\label{27}
\delta J(g)=\int_0^T \left [ \phi(\ell,t)+C^2_{\theta}\,\phi_x(\ell,t)\right ]\delta g(t) dt+\frac{1}{2}\int_0^\ell \left (\delta u(\ell,t) \right )^2 dt,\,g \in \mathcal{G}.
\end{eqnarray}

\medskip
\begin{theorem}\label{Theorem-3}
Assume that the inputs in (\ref{1-1}) satisfy the basic conditions (\ref{1-3}). Suppose, in addition, the measured output $\nu(t)$ belongs to $H^1(0,T)$. Then the Tikhonov functional introduced in (\ref{1-7}) is Fr\'{e}chet differentiable. Furthermore, for the Fr\'{e}chet gradient of this functional the following gradient formula holds:
\begin{eqnarray}\label{28}
\nabla J(g)(t)=\phi(\ell,t)+C^2_{\theta}\,\phi_x(\ell,t),\,t \in (0,T),\,g \in \mathcal{G}.
\end{eqnarray}
\end{theorem}
{\bf Proof.} Applying the first inequality in (\ref{13}) with the second estimate in (\ref{11}) to the weak solution $\delta u(x,t)$ of problem (\ref{18}) we conclude that the second right hand side integral in (\ref{27}) of the order $\mathcal{O}( \Vert g' \Vert_{L^2(0,T)})$. This means that the Tikhonov functional is Fr\'{e}chet differentiable.
 \hfill$\Box$

The gradient formula (\ref{28}) expressed in terms of the weak solution $\phi(x,t)$ of the adjoint problem (\ref{25}) forms the basis of the algorithm for numerical solving the inverse problem (\ref{1-1})-(\ref{1-2}).

\section{Numerical Algorithms and Computational Experiments}

In this section, a detailed description of an efficient numerical method is presented to solve the inverse problem (\ref{1-1})-(\ref{1-2}). This process has several steps and each of them should be considered carefully due to the sensitivity of the identification process.  First,  measured data $\nu(x):=u(\ell,t)$ is generated by solving the direct problem. It is critical to keep the error as low as possible in this step. This requires a successful algorithm for the solution of the direct problem (\ref{1-1}). Due to the effectiveness of the method of lines approach used in our several published previous studies (\cite{Hasanov-Baysal-Sebu:2019},\cite{AH:OB2015},\cite{AH:OB2016},\cite{Hasanov-Baysal-Itou:2019}) on an optimized mesh, an improved version of this method is employed here.

\subsection{The Method of Lines (MOL) Approach for the Numerical Solution of Direct Problem}

Basically, the MOL  is based on the principle of independent discretization of space and time variables. More specifically, a semi-analytical structure is obtained by expressing the variational formulation in finite dimensional space denoted by $V_h$. The method here is a finite element approximation with cubic Hermite basis functions which ensures continuity of both deflection and slope
throughout the beam. These shape functions is defined on uniformly discretizing spatial domain $0=x_0<x_1<\cdots<x_{M}=\ell$ (where $h=\ell/M$).
Formally, the solution $U_h(t):=u_h(\cdot,t)\approx u(\cdot,t)$ satisfies the following semi-discrete version of the variational formulation of (\ref{1-1}).

\emph{For all $t\in(0,T]$, find $U_h(t)\in  V_h \subset \mathcal{V}^2(0,\ell)$ such that $\forall v_h\in V_h$
\begin{eqnarray}\label{semidisc}
\left\{\begin{array}{ll}
(\rho_A(\cdot) U_h''(t),v_h)+(\mu(\cdot)U_h'(t),v_h)+a_r(U_h(t),v_h)+a_\kappa(U_h'(t),v_h) =0,\\
U_h(0)=0,~~ U_h'(0)=0.
\end{array}\right.
\end{eqnarray}}
Here the symmetric bilinear functional $a_\psi:H^2(0,\ell)\times H^2(0,\ell)\rightarrow \mathbb{R}$ is defined, for $\psi\in L^2(0,\ell)$, by $$a_\psi(u,v):=(\psi u_{xx},v_{xx}).$$
The next discretization step is performed for temporal derivatives. At this level  the second order system of ODE in (\ref{semidisc})  can be approximately solved by using any temporal finite difference method. It is crucial that the approach to be used here have to be practical, fast and stable.
These requirements can be met through the following second order backward finite difference approximations of $U_h''$ and $U_h'$ with uniform temporal discretization
$0=t_0<t_1<\cdots<t_{N}=T$ (where $\tau=T/N$).
 \begin{eqnarray}\label{imptimeint}
&&U_h''(t_j)\approx \partial_{\tau \tau}U_h^j:=\frac{2U_h^{j}-5U_h^{j-1}+4U_h^{j-2}-U_h^{j-3}}{\tau^2},\nonumber \\
&&U'(t_j)\approx \partial_{\tau}U_h^j:=\frac{3U_h^{j}-4U_h^{j-1}+U_h^{j-2}}{2\tau}, \nonumber
\end{eqnarray}
The full-discrete algebraic systems of equations are obtained by substituting these difference expressions with $U''_h(t)$ and $U'_h(t)$ in (\ref{semidisc}). Solutions of the resulted equations are provided desired approximations $U_h^j\approx u(x,t_j)$ for $j=0:N$. Note that for $j=1,2$, the necessary a priori approximations can be obtained by combining the ghost point technique within the central difference scheme.

Finally, several numerical tests are compared to determine the effective values of the pair $(M,N)$ and optimized with the ratio $h/\tau\simeq 142$.

\subsection{Reconstruction with Conjugate Gradient Algorithm (CGA)}

The explicit gradient formula in (\ref{28}) is very important in determining the minimizer of the least square functional (\ref{1-5}) for any unconstrained optimization techniques. Here we use CGA, one of the most suitable and stable one. It is known that this method is based on the conjugate directions and these directions are determined by the solution of the adjoint problem (\ref{25}). This requires the MOL technique at each iteration step. Although CGA is a self-stabilized method, the quality of the reconstruction process also depends on the success of solving both adjoint and direct problem. The details of the CGA is as follows.

\begin{itemize}
  \item From $g^{(i)}(t)$, calculate the decent direction
  \begin{eqnarray*}
			p^{(i)}(t)=
			\frac{\Vert  \nabla J(g^{(i)})\Vert^2_{L^2(0,T)}}{\Vert \nabla J(g^{(i-1)})\Vert^2_{L^2(0,T)}}~p^{(i-1)}(t)-\nabla J(g^{(i)})(t).
	\end{eqnarray*}
  \item Define the next iteration $g^{(i+1)}(t)=g^{(i)}(t)+\alpha_*^{(i)} p^{(i)}(t)$.
  Here $\alpha_i^*$ solution of the minimization problem
  $$J(g^{(i)}(t)+\alpha_*^{(i)} p^{(i)}(t))=\min_{\alpha >0}J(g^{(i)}(t)+\alpha p^{(i)}(t)); $$
  and has the following explicit form,
  $$ \displaystyle \alpha_*^{(i)} =\frac{||\nabla J(g^{(i)})||_{L^2(0,T)}^2}{|| u(\ell,\cdot,p^{(i)}) ||_{L^2(0,T)}^2} $$
  \item If the following stopping condition based on Mozorov’s discrepancy principle holds, $$ \Vert u(\ell,\cdot;g^{(i)})-\nu^{\gamma} \Vert_{L^2(0,T)}\leq \varepsilon\gamma <
     \Vert u(\ell,\cdot;g^{(i-1)})-\nu^{\gamma} \Vert_{L^2(0,T)} $$  for known parameter $\varepsilon >0$, stop the iteration; otherwise, repeat the process by taking $g^{(i)}(t):=g^{(i+1)}(t)$.
 \end{itemize}

For the first iteration, an arbitrary choice of $g^{(0)}(t)$  can be made, but if there is no prior knowledge it is better to choose $g^{(0)}(t)=0$ and $p^{(i)}(t):=-\nabla J(g^{(0)})(t)$. As a note, the first iteration has no significant effect to the success of the algorithm.

Here a standard method is used for the derivation of synthetic noise with a given noise level $\gamma>0$. In deed, the formula  $\nu^{\gamma}(t_j):=\nu(t_j)+ \gamma \, \Vert \nu\Vert_{L^2(0,T)}\, R_j$ for $j=1,\cdots, N$ generates measured noisy data. The  vector $R$ has $M$ random numbers array normally distributed with mean $0$ and standard deviation $\sigma=1$.

In CGA steps, both Fr\'{e}chet derivative $\nabla J(g)$ and $L^2(0,T)$ norms are computed by Simpson's numerical integration while in MOL algorithm, a  three-point Gauss quadrature rule is employed for all computation on each element.

\subsection{Computational Experiments}
In the reconstruction process, we work on two different test problems. One of them is based on engineering applications (realistic parameters), while the other one is preferred to test the applicability of the method.

It is a general approach to use error analysis when comparing the quality of the  methods. In the literature, two quantities are frequently used. These are Convergence  and Accuracy Errors as follows.
 $$\underbrace{e(i;g;\gamma)=\Vert\nu(\cdot;g^{(i)})-\nu^{\gamma}\,\Vert_{L^2(0,T)}}_{Convergence ~~Error}~~~{\rm{and}}~~~\underbrace{E(i;g;\gamma)=\Vert g-g^{(i)}\Vert_{L^2(0,T)}}_{Accuracy ~~ Error}.$$

As can be seen from their definitions, the Accuracy Error determines the success of the reconstruction. On the other hand, especially in the case of noisy data, the stop criterion is very crucial to prevent divergence of the approximation and it is completely related to Convergence Error. Therefore, these quantities should be evaluated together to analyze the process.

For \emph{the first test problem}, the parameters are selected in accordance with real engineering applications and are based on those proposed in  \cite{Turner-Wiehn:2001, McLaughlin:1994, Japanese:2009}. We take a beam of length of $200\, \mbox{nm}$ and observe it for a time interval of $10^{-3}\, \mbox{s}$. After a simple change of variables, to re-scale the problem so that the length of the beam  and the time observation length interval become  $\ell=1$ and  $T=1$ respectively, the numerical values adopted for this study become as follows:
\begin{eqnarray*}
\rho_A(x)=1.864\times 10^{-7} \, \mbox{kg/nm}, ~  \, \mu(x)=8.16 \times 10^{-6} \,
\mbox{kg \, s}^{-1}/\mbox{nm},\\
r(x)=2.265\times 10^{-3} \, \mbox{kg\,  nm}^3/\mbox{s}^{-2}, ~ \, \kappa(x)=3.5875\times 10^{-5} \, \mbox{kg\, nm}^3/\mbox{s},
\end{eqnarray*}
and domain parameters are $\ell=1$ and $T=1$, both non-dimensional. As for the tip length, it usually ranges from $5 \ \mbox{nm}$ to $50 \ \mbox{nm}$ \cite{AFM2012}. After the re-scaling for doing our numerical simulations, we take as a reasonable value $h=0.2$ (non-dimensional).

\begin{figure}[!htb]	\includegraphics[width=7.5cm,height=7.0cm]{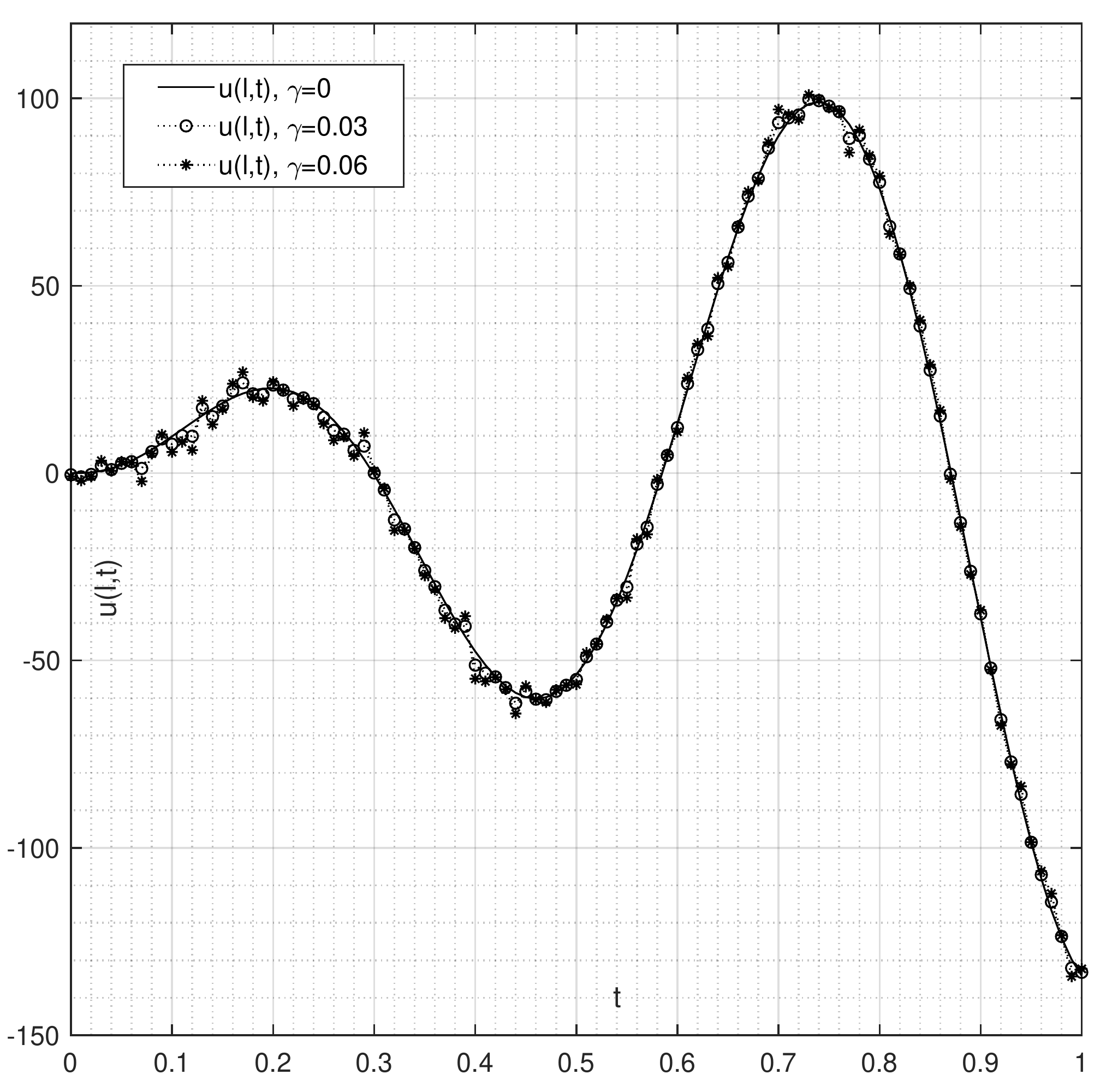}~~\includegraphics[width=7.0cm,height=7.5cm]{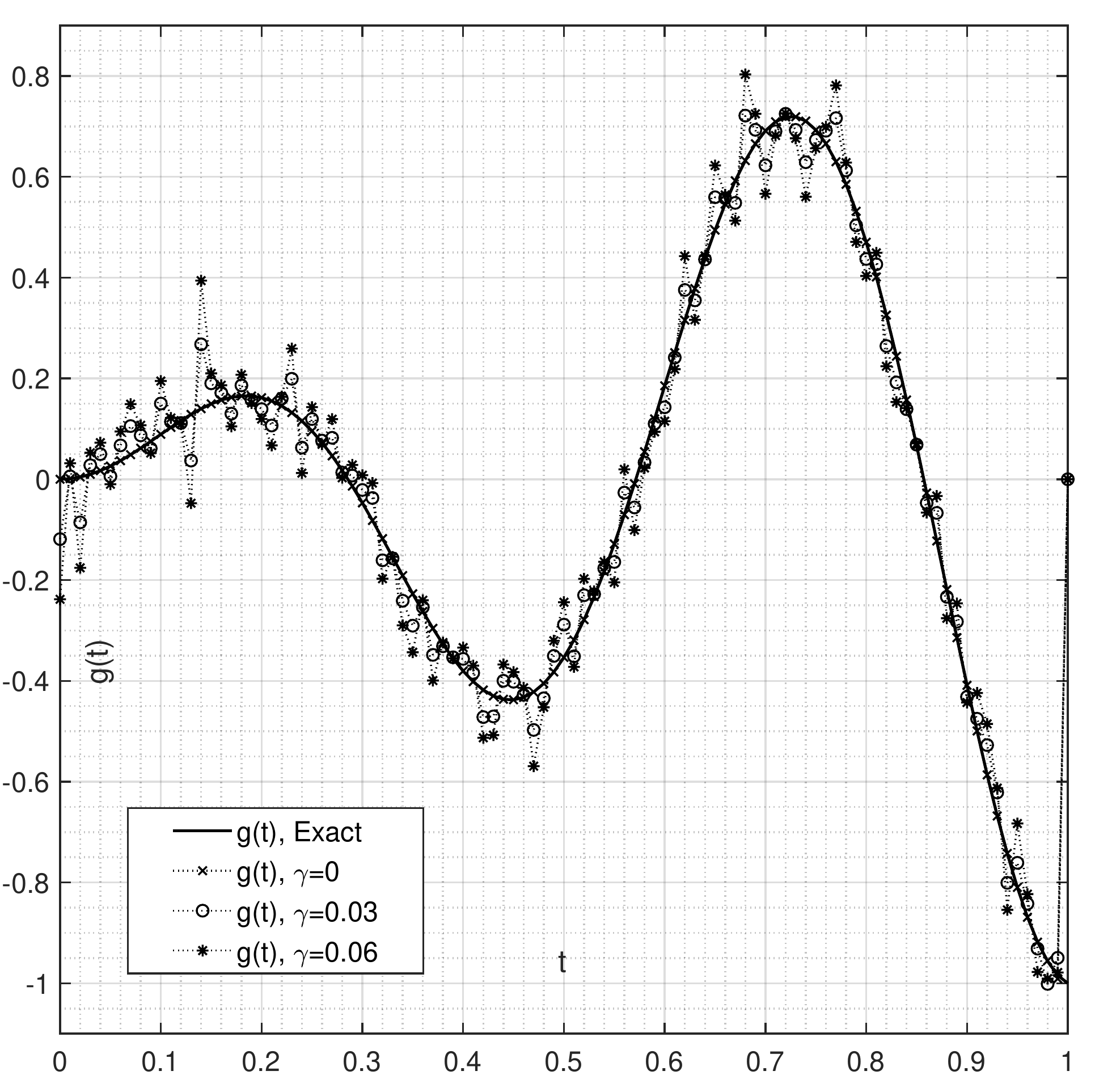}
	\caption{Synthetic noise free and noisy output data (left), reconstruction of smooth shear force $g(t)= t\ \sin(7\pi t/2)$ (right).} \label{Fig2a}
\end{figure}

We tested the performance of the algorithm for the unknown shear force  $g(t)= t\ \sin(7\pi t/2)$ with $\theta_\pi=\left (\cos(\pi/36)\right )/(5\pi)$.
The graph on the left in Fig. \ref{Fig2a} shows noisy free as well as random noisy output data with the noise levels $\gamma= 3\%$ and $6 \%$.
Then unknown target $g(t)$ is identified by using each of these data. Results can be seen on the right in Fig. \ref{Fig2a}.

\begin{figure}[!htb]
    \includegraphics[width=7.5cm,height=7.0cm]{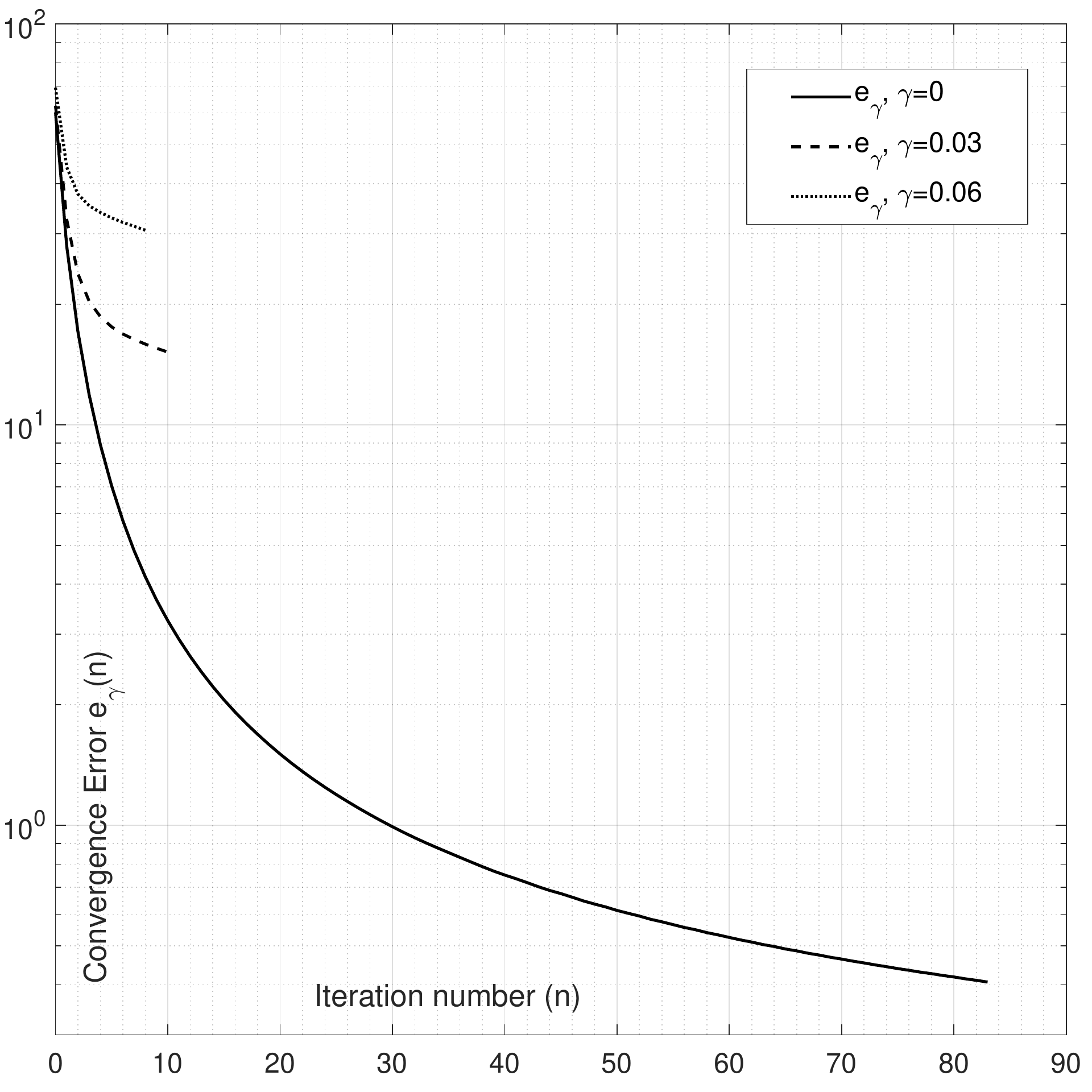}~~\includegraphics[width=7.0cm,height=7.5cm]{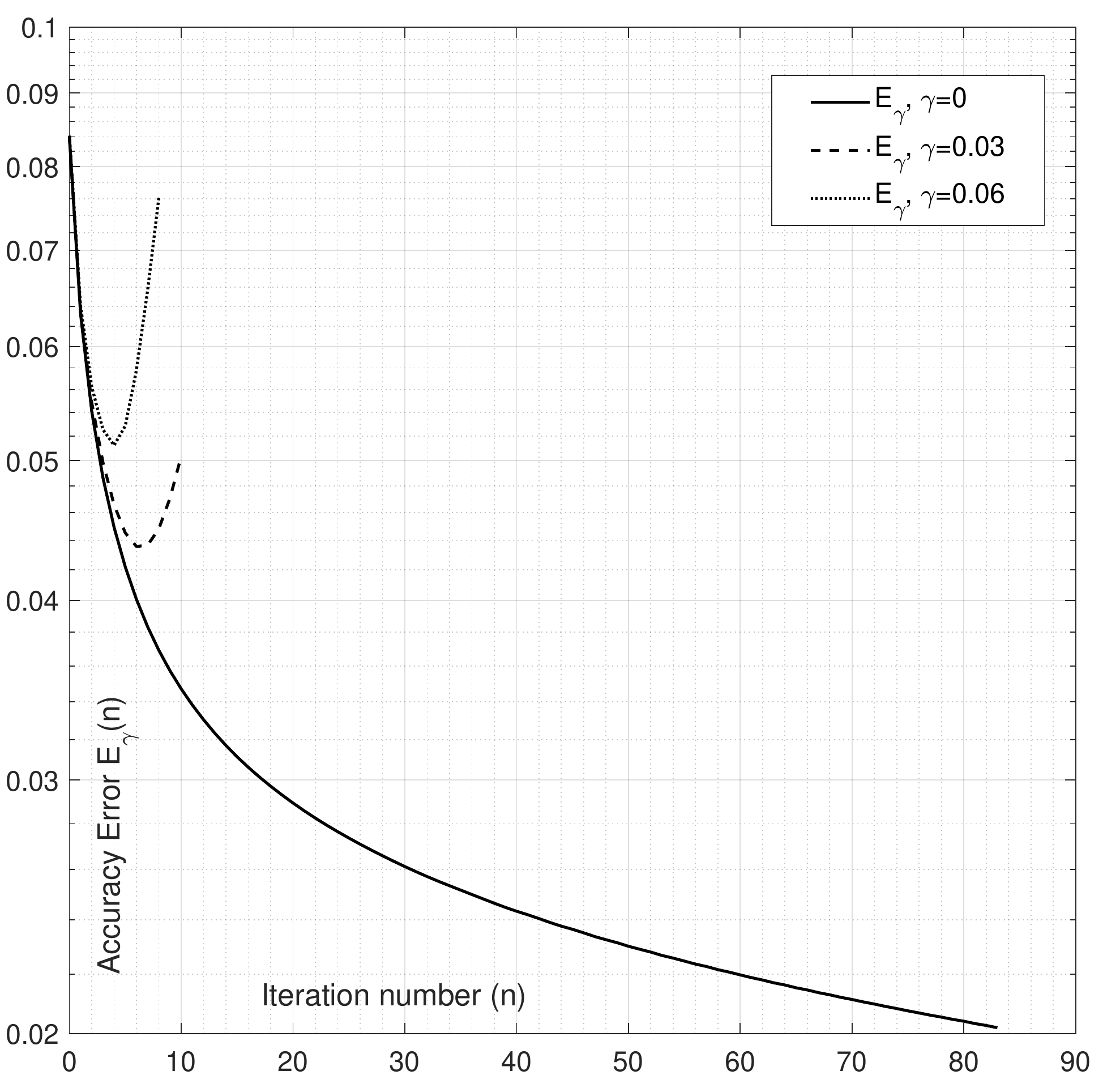}
	\caption{Convergence error (left) and accuracy error (right) for $g(t)= t\ \sin(7\pi t/2)$.} \label{Fig2b}
\end{figure}

Fig.\ref{Fig2b} reveals the general characteristics of an iteration. Especially the rapid deterioration in the Accuracy Error indicates that the sensitivity of the stopping which is directly determined by the Convergence Error. In case this balance is not determined appropriately, the success of the construction process can be adversely affected.

\begin{figure}[!htb]
	\includegraphics[width=7.0cm,height=7.0cm]{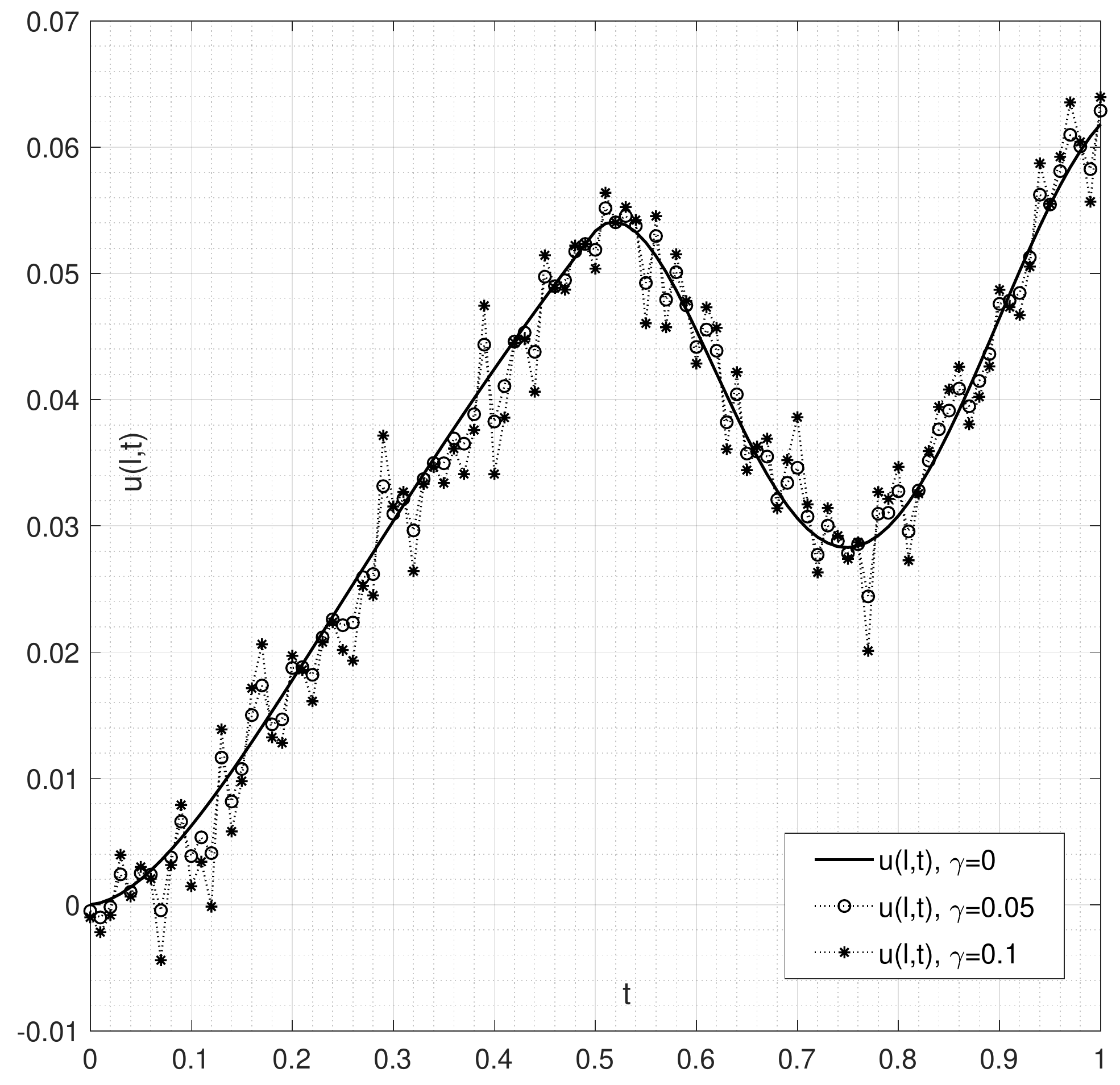}~~\includegraphics[width=7.0cm,height=7.0cm]{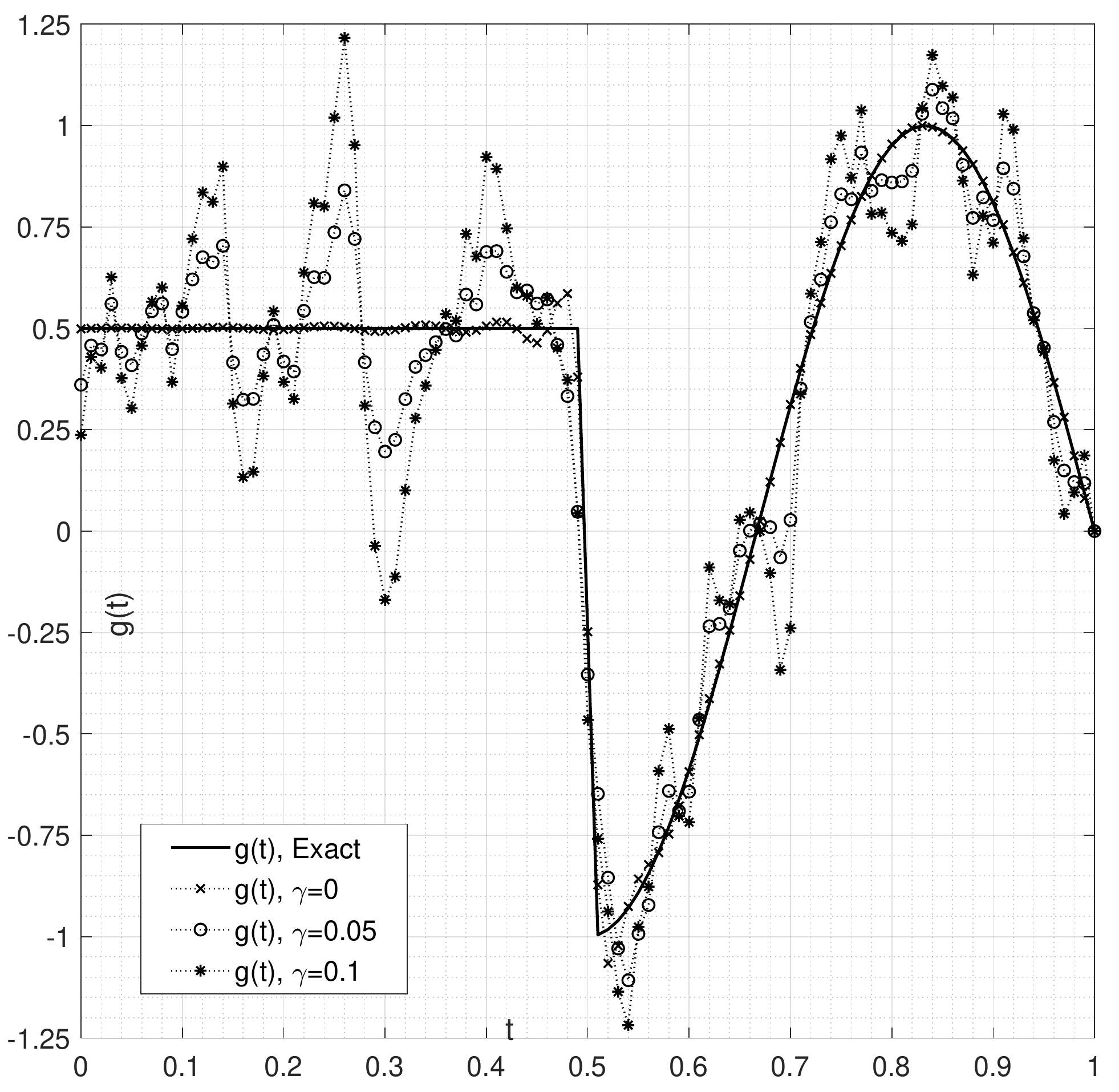}
	\caption{Synthetic noise free and noisy output data (left), reconstruction of non-smooth shear force $g(t)=\frac{1}{2}H(1/2-t)+ \sin(3\pi t)\cdot H(t-1/2)$ (right).} \label{Fig3a}
\end{figure}

\emph{The second computational experiment} aims to test the accuracy of CGA regardless of the realizability of the parameters. For this goal, reconstruction of the following discontinues target source $g(t)$ is studied  under high noise levels.
$$g(t)=\frac{1}{2}H(1/2-t)+ \sin(3\pi t)\cdot H(t-1/2) ~~{\rm{and}}~~ \displaystyle \theta_\pi=\frac{2\cos(\pi/4)}{\pi}$$
Here $H(x)$ is the Heaviside step function. Moreover, all problem parameters are imposed as non-constant case as follows with unit domain parameters $\ell=1$ and $T=1$.
$$ \rho_A(x)=\exp(x), ~~ \mu(x)=\sin(\pi x),~~ r(x)=2+x^2,~~ \kappa(x)=1+\exp(-x)$$
Synthetic noise free and noisy data are plotted in Fig. \ref{Fig3a} (left)  with  noise levels $\gamma= 5\%$ and $10 \%$. Then CGA is applied for identification of the temporal function $g(t)$ and results are illustrated in Fig. \ref{Fig3a} (right). Here, due to the effect of high noise levels and discontinuity on $g(t)$, non-physical distortions are naturally observed in the reconstruction.

\begin{figure}[!htb]
    \includegraphics[width=7.5cm,height=7.0cm]{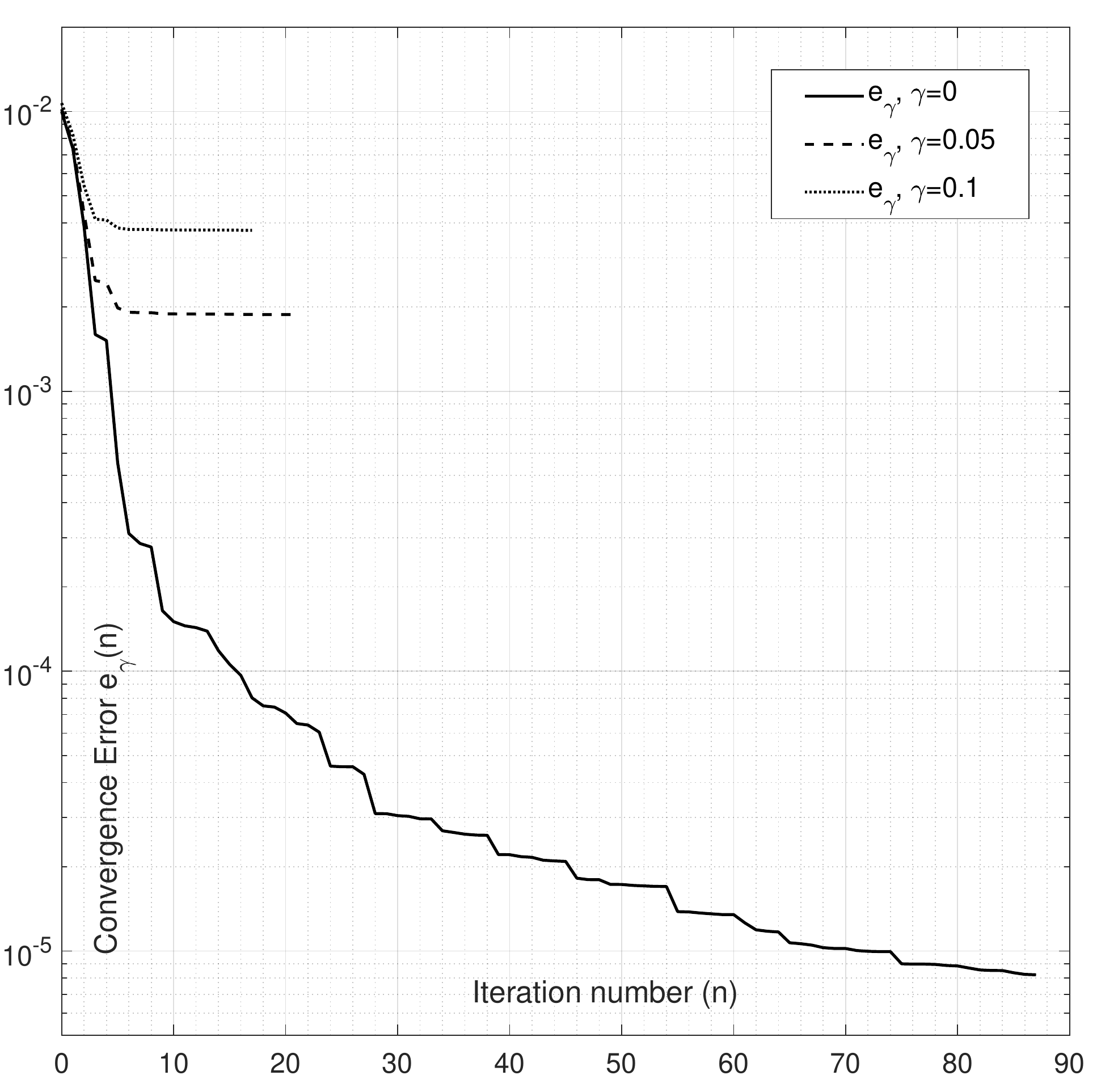}~~\includegraphics[width=7.0cm,height=7.5cm]{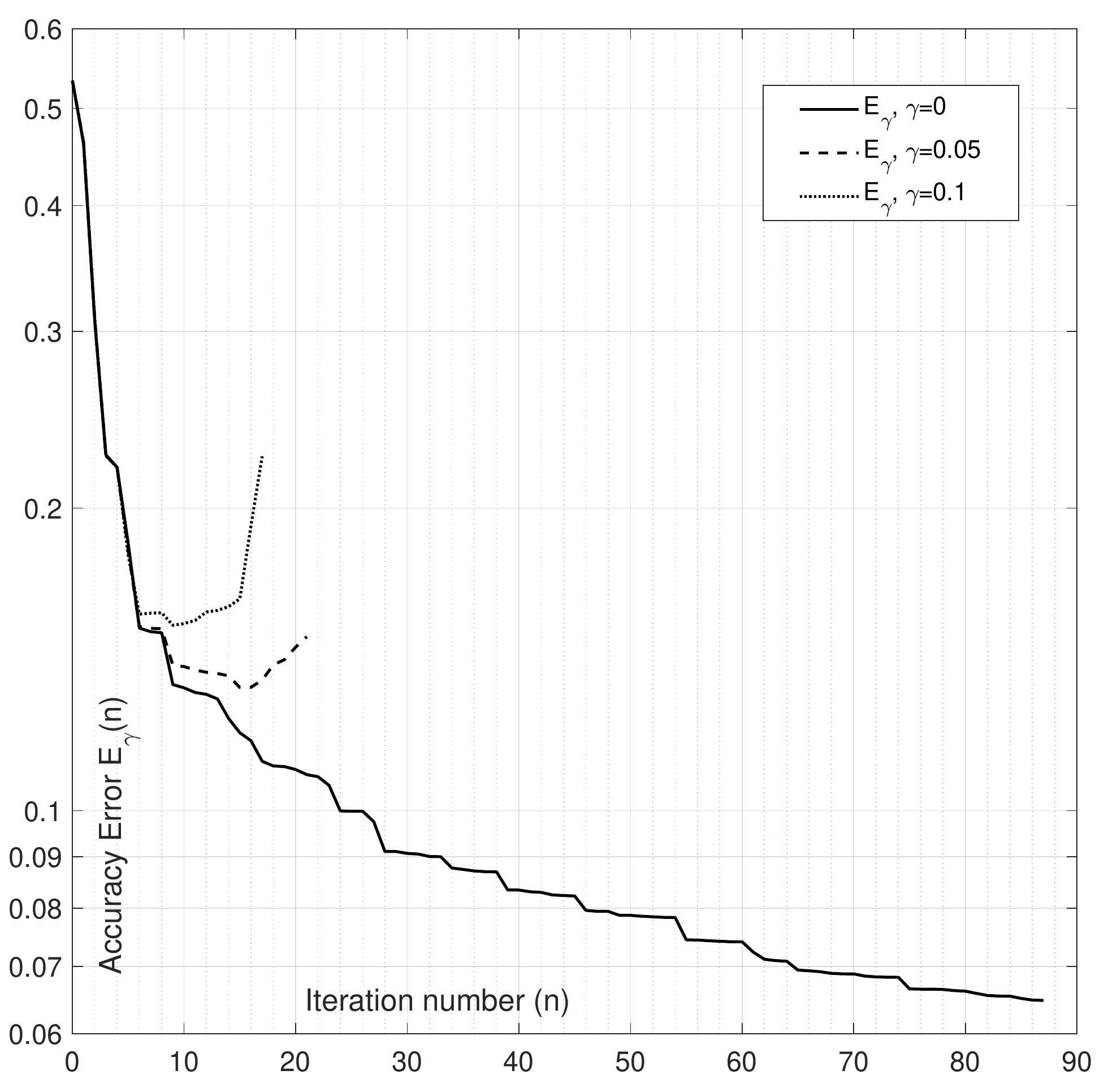}
	\caption{Convergence error (left) and accuracy error (right) for $g(t)=\frac{1}{2}H(1/2-t)+ \sin(3\pi t)\cdot H(t-1/2)$.} \label{Fig3b}
\end{figure}

Convergence and Accuracy Errors are plotted in Figure \ref{Fig3b} on the left and on the right, respectively. Similar behavior of these error quantities examined in the first problem is also observed in this second experiment.

The results of the two experiments presented here show that CGA is  effective and successful for the solution of the inverse problem under consideration, provided that certain sensitivities are taken into account. Nevertheless, the algorithm may need to be improved for further applications. Especially in the realistic cases, it is required to choose a small final time such as $T=10^{-3}$ for stable calculations using the Finite Element Method. Since the method suggested here is just a preliminary numerical study of the inverse problem related to Atomic Force Microscopy, we only aimed to present the general principles.

\section{Conclusions}

In this study, a novel mathematical model of tip-sample processing with AFM cone-shaped cantilever is proposed. Compared to the models known in the literature, this model is a fairly advanced model, and takes into account both viscous and internal damping parameters. A detailed mathematical analysis of the model has been carried out. An explicit gradient formula for the Fr\'echet derivative of Tikhonov functional is derived through the weak solution of the appropriate adjoint problem. This allows us to construct the fast Conjugate Gradient Algorithm for the numerical reconstruction of the shear force. Numerical experiments carried out with real physical and geometric parameters show the high accuracy of the algorithm.

\section*{Acknowledgments}

The research of the first and second authors have been supported by FAPESP, through the Visiting Researcher Program, proc. 2021/08936-1, in Escola Polit\'{e}cnica, University of S\~{a}o Paulo, Brazil, during the period November 02 - December 18, 2022.

\end{document}